\documentclass[11pt]{article}
\usepackage{jcappub}
\usepackage{array}

\usepackage{graphicx}% Include figure files
\usepackage{epsfig}

\newcommand{\be}{\begin{equation}}
\newcommand{\ee}{\end{equation}}

\newcommand{\hi}{\mbox{H\,{\scriptsize I}\ }}

\title{A joint analysis of Planck and BICEP2 B modes including dust polarization uncertainty}

\author[a]{Michael J.\ Mortonson,} \emailAdd{mmortonson@berkeley.edu}
\author[b]{Uro\v s Seljak} \emailAdd{useljak@berkeley.edu}

\affiliation[a]{Space Sciences Laboratory and Lawrence Berkeley National Laboratory,
University of California, Berkeley CA 94720, USA}

\affiliation[b]{Department of Physics, Department of Astronomy,
and Lawrence Berkeley National Laboratory, University of California, Berkeley CA 94720, USA}

\abstract{
We analyze BICEP2 and Planck data using a model that includes CMB lensing, gravity waves, and 
polarized dust. 
Recently published Planck dust polarization maps have 
highlighted the difficulty of estimating the amount of
dust polarization in low intensity regions, 
suggesting that the polarization fractions have considerable uncertainties and may be
significantly higher than previous predictions. In this paper,
we start by assuming nothing about the dust polarization except for the power
spectrum shape, which 
we take to be $C_{l}^{BB,{\rm dust}} \propto l^{-2.42}$. 
The resulting joint BICEP2+Planck analysis favors solutions 
without gravity waves, and the upper limit on the tensor-to-scalar ratio is $r<0.11$, a slight improvement
relative to the Planck analysis alone which gives $r<0.13$ (95\%\,c.l.). 
The estimated amplitude of the dust polarization power spectrum agrees 
with expectations for this field based on both \hi column density 
and Planck polarization measurements at 353\,GHz in the BICEP2 field. 
Including the latter constraint on the dust spectrum amplitude 
in our analysis improves the limit further to $r < 0.09$, placing 
strong constraints on theories of inflation (e.g., models with $r>0.14$
are excluded with 99.5\% confidence).
We address the cross-correlation analysis of BICEP2 at 150\,GHz 
with BICEP1 at 100\,GHz as a test of 
foreground contamination. We find that the null hypothesis of dust and lensing 
with $r=0$ gives 
$\Delta \chi^2<2$ relative to the hypothesis of no dust, so the frequency
analysis does not strongly favor either model over the other. 
We also discuss how more accurate 
dust polarization maps may improve our constraints. 
If the dust polarization is measured perfectly, the limit can reach $r<0.05$ (or the corresponding 
detection significance if the observed dust signal plus the expected lensing signal
is below the BICEP2 observations), but this degrades quickly 
to almost no improvement if the dust calibration error is 20\% or larger or if the dust 
maps are not processed through the BICEP2 pipeline, inducing sampling variance noise. 
}

\begin{document}
\maketitle
\flushbottom

\section{Introduction}

$B$ mode polarization of cosmic microwave background (CMB) anisotropies 
is a powerful experimental approach to study inflationary
models and is expected to improve upon the constraints from the 
temperature anisotropies (see \cite{2013arXiv1303.5082P} for a review). This is because $B$ modes
can be generated by the primordial gravity waves from inflation \cite{1997ApJ...482....6S,1997PhRvD..55.1830Z,1997PhRvD..55.7368K} and not by ordinary scalar modes, meaning 
that the signal can be detected without the sampling variance from scalar
perturbations which dominates other CMB 
channels such as temperature and $E$ polarization. 
The above statement needs to be 
modified to some extent, since gravitational lensing can generate $B$ polarization by lensing $E$ polarization 
\cite{1998PhRvD..58b3003Z}. 
At high $l$ this lensing signal dominates, and the small gravity wave signal
is masked by the sampling variance uncertainties. Only for
$l<100$ is the lensing signal sufficiently small to be able to probe gravity waves below the 
current limits set by the Planck satellite \cite{2013arXiv1303.5082P}. Nevertheless, one expects that $B$ modes will 
dominate the limits or detections of primordial gravity waves in the future. 

At the noise sensitivity level this expectation has already been reached by the BICEP2 experiment: 
their recent results \cite{2014arXiv1403.3985B} have generated a great deal of excitement in the field, with an excess 
signal seen above the expectations from lensing. Original estimates of the foreground contribution, primarily 
dust polarization, suggested that it was at most a minor correction to the observed signal, of order 25\%. However,  
recently released dust polarization maps from the Planck satellite suggest a relatively high 
polarization fraction in low intensity regions of the sky, around 8--10\% on
average but with a lot 
of scatter \cite{2014arXiv1405.0871P}. These maps do not include the BICEP2
region,\footnote{During the late stages of the review process for this paper,
an analysis of Planck polarization maps that include the BICEP2 region was 
presented in \cite{2014arXiv1409.5738P}. We briefly discuss the impact of 
these new results in later sections.}
citing as some of the reasons both noise and 
residual systematic uncertainties, especially contamination from the cosmic infrared background
(CIB). The BICEP2 team assumed a polarization fraction 
of 5\% for dust in their field, based on a preliminary map presented at a conference \cite{Bernard}. 
A visual comparison of this 
map with the new version in \cite{2014arXiv1405.0871P} suggests that there is imperfect
agreement between the two in many regions 
and that the polarization fractions are significantly higher in the new maps
relative to the old ones. One reason for the discrepancy is the CIB, which was not 
corrected for in the old maps; since CIB is not polarized correcting for it reduces intensity but not 
polarization, increasing the polarization fraction. 
We note that changing the polarization fraction from 5\% to 8.5\%
would increase the dust polarization power spectrum by a factor of 3, which would result in a very large 
signal, potentially able to explain most of the observed signal as dust. 
In addition, the ``DDM2'' dust polarization estimate presented by the BICEP2 team, which is the most 
realistic model they consider, lacks degree-scale angular variations of the polarization 
direction, making it uncertain at higher $l$.
(The ``DDM1'' model lacks fluctuations in the polarization fraction as well, making it even more of an underestimate). 
Given the uncertainties in 
the CIB, including the monopole, the estimates of the intensity zero modes, etc., 
the dust power could be significantly underestimated. Furthermore,  
because the degree-scale angles are unreliable, it is perhaps unsurprising that the
cross-correlation of BICEP2 data with such an imperfect map results 
in little or no correlation, making the cross-correlation an unreliable test of dust foregrounds. 
Overall, using the polarization fraction from 
a preliminary map, without a proper error analysis or understanding of 
its systematics, is not reliable, 
even more so now that we know that the map has changed considerably in the 
new version published in \cite{2014arXiv1405.0871P}.\footnote{Note that in the
peer-reviewed, published revision of the BICEP2 paper 
\cite{2014PhRvL.112x1101A}, which appeared after the preprint of our paper,
the BICEP2 team acknowledges many of these issues, including 
the larger uncertainty in the level of dust contamination suggested by 
the new measurements from \cite{2014arXiv1405.0871P}; 
their revised paper also omits the ``DDM2'' dust polarization model.}

Given all these considerations, in this paper we take a step back, ignore any information about the 
dust polarization that may or may not be currently available in the BICEP2 field
and revisit the analysis using a more conservative approach. 
We ask the question: what limits on $r$ can we deduce from BICEP2 and Planck given what 
we currently know about dust foregrounds?
One of the more robust results presented by the Planck team \cite{Aumont} 
is that the power spectrum of the dust polarization
scales as $\Delta_{BB,{\rm dust},l}^2=l^2C_{l}^{BB,{\rm dust}}/2\pi
\propto l^{-0.3}$ independent of the amount of dust, 
with an overall amplitude that strongly depends on the amount of dust 
in the field. Recent Planck analyses confirm the scalings presented in \cite{Aumont} for dust 
intensity \cite{2014arXiv1405.0874P} 
and polarization \cite{2014arXiv1409.5738P}. 
For the BICEP2 dust column density, which based on the dust maps is estimated to be 
$N_{\hi}=1.5\times 10^{20}\,{\rm cm}^{-2}$ \cite{Flauger}, extrapolation using the scaling
presented in \cite{Aumont} 
suggests $\Delta_{BB,{\rm dust},100}^2 \sim 0.015\,\mu{\rm K}^2$, a factor of 5 higher
than the BICEP2 DDM2 estimate. 
Although this extrapolation is rather uncertain, it once again 
suggests that one should worry about the overall dust polarization 
levels. In this paper, we will retain only the power-law scaling of the
dust spectrum with $\Delta_{BB,{\rm dust},l}^2 \propto
l^{-0.42}$,\footnote{This scaling matches the best estimate of the power 
spectrum shape from recent Planck data over large regions of the sky 
\cite{2014arXiv1409.5738P}. We have also performed our analysis with the
$l^{-0.3}$ scaling from \cite{Aumont}, and find 
that this small change in the exponent has little impact on our results,
affecting the upper limits on $r$ at the level of a few percent.}
assuming a flat amplitude prior between 0 and 
$\Delta_{BB,{\rm dust},100}^2=0.03\,\mu{\rm K}^2$.
We will also assume that the covariance matrix for the dust polarization component is Gaussian; this is the 
most optimistic assumption, as there may be additional non-Gaussian bandpower correlations present which we will neglect here. 

Unless otherwise specified, throughout this paper $r$ will denote the tensor-to-scalar ratio at 
$k=0.05\,{\rm Mpc}^{-1}$, upper limits are reported at 95\%\,c.l., 
and constraints with both upper and lower limits are 68\%\,c.l.\ ranges.

\section{Likelihood analysis}

We evaluate combined constraints from BICEP2 and Planck data 
by importance sampling Planck parameter chains using the BICEP2 likelihood code.
Specifically, we use the ``Planck+WP'' chains from the 
Planck Legacy Archive (PLA),\footnote{\url{http://www.sciops.esa.int/index.php?page=Planck_Legacy_Archive&project=planck}} which 
include WMAP polarization constraints on the reionization optical depth in addition
to Planck temperature data, analyzed assuming the ``\verb+base_r+''
flat $\Lambda$CDM$+r$ model.
To speed up computations for some tests, we ``thin'' the chains by a factor of 
two, using every other model from the PLA chains in our analysis; 
we have verified that
this does not significantly change the constraint on $r$ 
from Planck+WP data alone ($r<0.13$).
Note that we only use the Planck team's analysis 
without high $l$ experiments (ACT and SPT), hence our limits on $r$ are somewhat weaker than those presented in \cite{2013arXiv1303.5076P}
and somewhat stronger than the Planck reanalysis presented in \cite{2013arXiv1312.3313S}. 

We consider two analyses. The main one for the purpose of deriving current 
constraints assumes there is sampling variance on the 
dust polarization amplitude in the BICEP2 observing field, as expected if the 
amplitude prior is based on observations over a larger region of 
the sky. We then also ask how much better one will be able to do 
with future data, such as Planck, 
if the dust prior
is based on measurements in the BICEP2 field and therefore has 
no sampling variance, but may still have some overall error that could
arise from noise or systematics.

For the first analysis, we use CAMB \cite{2000ApJ...538..473L} to compute the predicted $B$ mode power spectrum,
including both lensing and gravity wave components, 
for each sample in the thinned PLA chains. Assuming that the dust foreground
scales as $\Delta_{BB,{\rm dust},l}^2 \propto l^{-0.42}$ 
but that its amplitude is uncertain, we draw a random dust 
polarization amplitude from a flat prior, 
$0 \leq \Delta_{BB,{\rm dust},100}^2 \leq 0.03\,\mu{\rm K}^2$, and add 
the dust spectrum with that amplitude to the $B$ mode spectrum.
To ensure that the dust amplitude prior is sampled reasonably well, we draw 
several amplitudes for each model in the chains.
We use the public BICEP2 likelihood code,\footnote{\url{http://bicepkeck.org/}}
including the 9 bandpowers from $l \sim 45$ to $l \sim 300$,
to evaluate the likelihood of the total lensing+gravity wave+dust $B$ mode
spectrum. We then use this likelihood to importance sample the PLA chains 
(see, e.g., Appendix B of \cite{2002PhRvD..66j3511L}) by multiplying the 
original weight of each model by the BICEP2 likelihood; the new weights
can then be used to compute marginalized parameter estimates that 
include constraints from both Planck and BICEP2.
While for most of the analyses we use all 9 bandpowers of BICEP2, 
we also explore the effects
of only using the first 5 bandpowers. 

In the second case, where the dust spectrum in the BICEP2 field is 
assumed to be known without sampling variance, we subtract the 
dust spectrum from the data points rather than adding it to each model.
To do so, we have to assume a specific realization of the dust 
spectrum in the BICEP2 field, which in turn determines what 
level of gravity waves will best fit the data. 
The objective in this case is to determine how uncertainty in the 
measured dust polarization spectrum propagates into uncertainty in $r$. 
For the purposes of this test, we assume that the BICEP2 measurements
are equal to the sum of a lensing $B$ mode spectrum, a gravity 
wave component with a particular tensor-to-scalar ratio,\footnote{Specifically, 
for lensing we use the expected bandpowers from 
\url{http://bicepkeck.org/B2_3yr_cl_expected_lensed_20140314.txt}, 
and for gravity waves we rescale the $r=0.1$ spectrum from 
\url{http://bicepkeck.org/B2_3yr_cl_expected_withB_20140314.txt}.}
a dust polarization spectrum, and instrumental noise bias ($N_l$), 
and that the same dust spectrum
is independently measured in the BICEP2 field with some uncertainty
in its normalization at 150\,GHz:
\begin{eqnarray}
\hat{\Delta}_{BB,{\rm BICEP2},l}^2 &=& \Delta_{BB,{\rm lens},l}^2 + \Delta_{BB,{\rm GW},l}^2
+ \Delta_{BB,{\rm dust},l}^2 + N_l \,, \\
\hat{\Delta}_{BB,{\rm dust},l}^2 &=& (1+\epsilon_{l}) \,
\Delta_{BB,{\rm dust},l}^2 \,, \nonumber
\end{eqnarray}
where hats indicate observed quantities. 
In the first 4 bandpowers ($l\leq 160$), 
we take the error on the dust spectrum to be a constant 
$\epsilon$ for each model, 
drawn from a Gaussian with mean zero and width $\sigma_{\rm dust}$.
To avoid getting artificially strong constraints related to the excess
power above the lensing spectrum in BICEP2 bandpowers 5--7, 
we set $\epsilon_{l}=0$ in the 5 highest-$l$ bandpowers.
We then study the expected upper limits or constraints 
on $r$ as a function of the 
uncertainty in the dust polarization amplitude, $\sigma_{\rm dust}$, 
by drawing several values of $\epsilon$ for each sample
in the PLA $\Lambda$CDM$+r$ chains, 
subtracting the resulting $\hat{\Delta}_{BB,{\rm dust},l}^2$
from the BICEP2 bandpowers, and importance sampling the chains using
the BICEP2 likelihood as described above. 
We note that of course we do not know what the 
actual realization of dust polarization will be, so our analysis 
is only meant to give an approximate idea of what one can expect 
from external dust polarization maps and at what level the 
results depend on residual noise and systematics in these maps. 

Since the PLA chains contain relatively few samples with $r\sim 0.2$, 
importance sampling is not expected to be accurate for such large 
values of $r$ and combined constraints with large $r$ should be 
interpreted with caution. However, we find that most of our analyses
limit $r$ to values $\lesssim 0.1$ where importance sampling should
be more reliable.

\section{Impact of polarized dust on BICEP2 inflation constraints}

\begin{figure}[t]
\begin{centering}
\includegraphics[width=5.in]{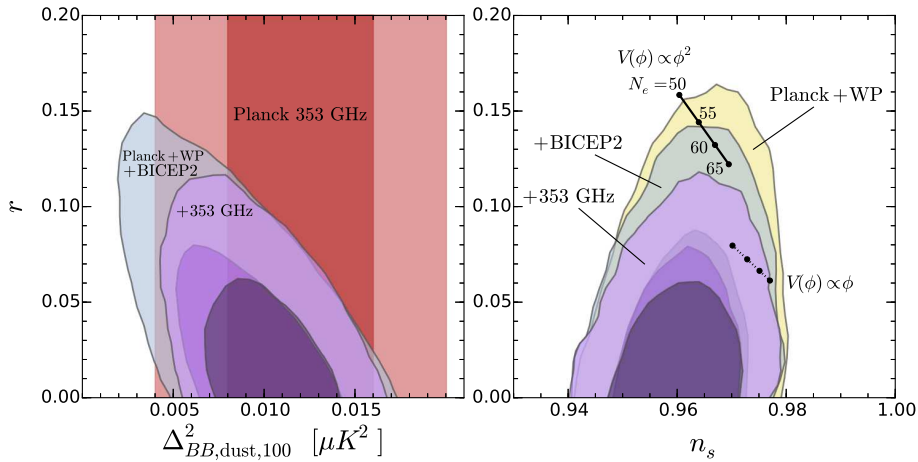}
\caption{\label{fig_2d}
\small Left: Joint constraints (68\% and 95\% c.l.) 
on $r$ and the amplitude of the 
dust polarization spectrum at $l=100$ from Planck+WP+BICEP2, 
assuming a flat prior on the dust amplitude (blue contours) or 
including the constraint on the dust polarization power in the BICEP2 field
estimated by extrapolating Planck data at 353\,GHz to 150\,GHz 
(red bands and purple contours).
Right: Constraints from the same combinations of data in 
the $r$--$n_s$ plane (blue and purple contours), compared with constraints
from Planck+WP alone (yellow contours). 
The thick solid line shows the relation between 
$n_s$ and $r$ predicted by inflation models with $\phi^2$ potentials
and the number of $e$-folds varying from 50 to 65; the dotted line 
shows the same relation for linear potentials.
}
\end{centering}
\end{figure}

We start with the most conservative analysis, where we assume
only a weak prior on the possible range of dust polarization 
amplitudes and no specific knowledge of the dust polarization in 
the BICEP2 field. The left panel of Figure~\ref{fig_2d} shows that
there is a clear anticorrelation in the 
resulting constraints 
on $r$ and the dust polarization amplitude $\Delta_{BB,{\rm dust},100}^2$. 
The joint constraints
favor models with small $r$ and $\Delta_{BB,{\rm dust},100}^2 \approx
0.01\,\mu{\rm K}^2$. This amplitude is 
consistent with the information about dust polarization 
in the BICEP2 field that is currently available.

At $2\,\sigma$ in this 2D parameter space, the contours 
do not extend to dust-free $r\approx 0.2$ models. In principle, the
preference for small $r$ could be driven by the Planck+WP constraints
which by themselves disfavor $r=0.2$ at almost $3\,\sigma$. 
However, we find
that even when considering the BICEP2 likelihood alone, models with $r=0$
and a polarized dust component fit the BICEP2 data better than 
models with $r=0.2$. In fact, despite marginalizing over the 
dust polarization amplitude, the joint constraint from Planck+WP+BICEP2
still imposes a slightly stronger limit on the tensor-to-scalar ratio ($r<0.11$)
than Planck+WP alone ($r<0.13$), if we use the first 9 bandpowers of BICEP2.
If we only use the first 5 bandpowers then the constraints are unchanged relative to Planck. 

The Planck collaboration recently released a new analysis of their data
that included estimates of 150\,GHz dust polarization power spectra 
in the BICEP2 field, inferred by extrapolating from data 
at 353\,GHz \cite{2014arXiv1409.5738P}.
These results show that the amplitude of B mode power from dust is
$(0.013 \pm 0.004)\,\mu{\rm K}^2$ in the band $40<l<120$, where the 
error includes noise and uncertainty from frequency extrapolation, 
added in quadrature. A full assessment of the impact of these new 
measurements on the interpretation of the BICEP2 data will have to wait
until the completion of the joint analysis that the BICEP2 and Planck 
teams are currently carrying out, but for now we can estimate the 
approximate effect on the constraint on $r$ (see also 
\cite{2014arXiv1409.7025C} which analyzes the same combination of data
and finds similar constraints).
We do this by retaining
our assumption of a power-law dust polarization spectrum with index
$-0.42$ (consistent with the measurements of \cite{2014arXiv1409.5738P} 
over larger regions of the sky) and simply 
replacing our assumption of a flat prior on the dust amplitude at $l=100$ with
a Gaussian constraint with mean $0.012\,\mu{\rm K}^2$ (adjusted downwards from 
the power at the midpoint of the measured band, $l=80$, using the 
power law scaling) and width $0.004\,\mu{\rm K}^2$. 
We still include sampling variance on the dust contribution in this case
since the shape of the dust spectrum in the BICEP2 field is poorly 
constrained. This is a somewhat conservative choice since the overall
amplitude of the spectrum in the BICEP2 field is measured without
sampling variance, but the extra variance is negligible relative to the 
$\sim 30\%$ uncertainty in the measured amplitude.
The left panel of Figure~\ref{fig_2d} shows that this constraint is fully
consistent with our initial result that assumed a flat prior on the dust
spectrum amplitude. Since the new Planck constraint favors models with 
relatively large contributions from dust, combining the constraints 
tightens the limit on the gravity wave component to $r < 0.09$.

In the right panel of Figure~\ref{fig_2d}, we show the constraints 
in the $r$--$n_s$ plane from this analysis compared with the constraints
from Planck+WP alone. While the upper limit on $r$ only improves 
slightly with the addition of BICEP2 data fit with a polarized 
dust component, the joint constraints place increasing pressure on 
large-$r$ inflation models such as quadratic potentials. This is even more 
true with the addition of the dust amplitude constraint from 
Planck 353\,GHz data, which excludes quadratic inflation models 
with $\sim 60$ $e$-folds of inflation at more than 
$2\,\sigma$ in the $r$--$n_s$ plane.

\begin{figure}[t]
\begin{centering}
\includegraphics[width=3.5in]{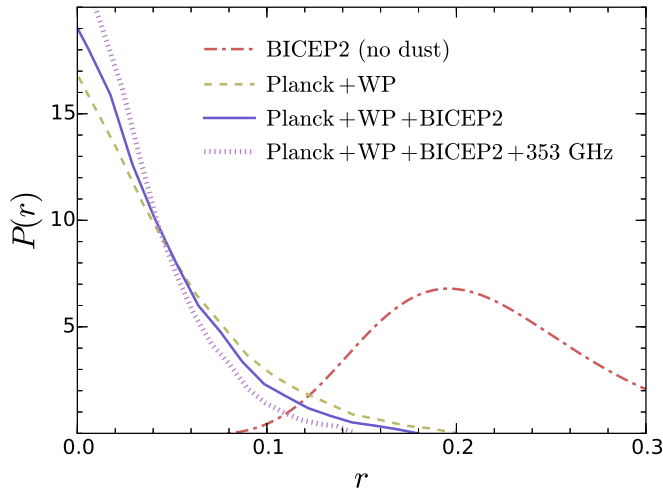}
\caption{\label{fig_1d}
\small Marginalized constraints on $r$ from Planck+WP (dashed curve), 
Planck+WP+BICEP2 with free dust polarization amplitude or 
including the constraint from Planck 353\,GHz data
(solid and dotted curves), and the BICEP2 likelihood alone (dot-dashed curve).
}
\end{centering}
\end{figure}

The 1D projection of the posterior probability for $r$ is 
shown in Figure~\ref{fig_1d} for both Planck+WP and 
Planck+WP+BICEP2, marginalized over the dust polarization amplitude 
with either a flat prior or the Planck 353\,GHz constraint.
We also plot the BICEP2 likelihood;
this is only intended as a qualitative comparison since, as noted
by \cite{2014arXiv1405.1390A}, the likelihood analysis presented in \cite{2014arXiv1403.3985B}
differs in several ways from the public likelihood 
code, leading to small shifts in $r$.
Despite these caveats, it is clear that the combined constraint 
from Planck+WP+BICEP2, allowing a free dust polarization amplitude,
is completely different from the main constraint from BICEP2 alone
which did not include a dust component.
We see that the combined analysis improves the limits at high values of $r$. 
For example, taking the prediction of a $V=M^2\phi^2/2$ model with 
about 55 $e$-folds of inflation, $r=0.14$, as a representative 
example, we find that only 1.7\% of the posterior distribution 
has $r>0.14$ in the combined analysis (and only 0.5\% with the Planck 353\,GHz
constraint), compared to 3.3\% for Planck+WP alone. 

\begin{figure}[t]
\begin{centering}
\includegraphics[width=3.8in]{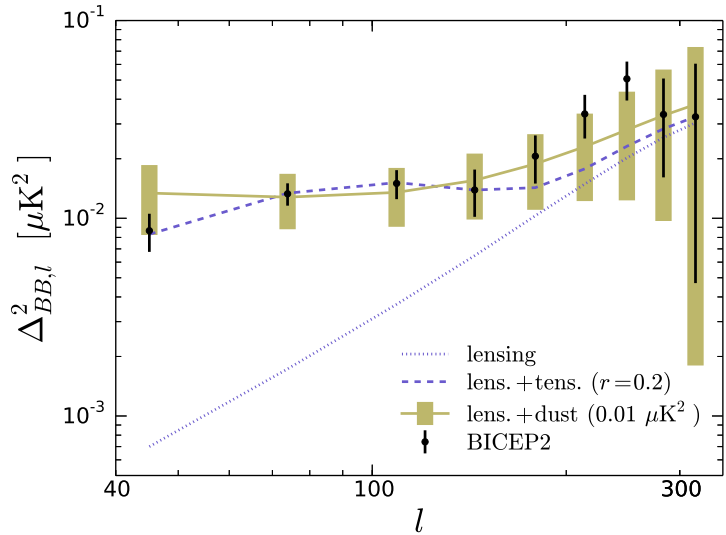}
\caption{\label{fig_bb}
\small $B$ mode spectrum predictions compared with BICEP2 data
(black points with error bars, showing measurement uncertainty only).
Each model curve shows the expected signal in the 9 BICEP2 bandpowers.
The dashed curve is the sum of a gravity wave component with $r=0.2$ and 
the lensing spectrum (dotted curve); the solid curve assumes 
$r=0$ and adds to the lensing spectrum a dust polarization spectrum
$\Delta_{BB,{\rm dust},l}^2 = (0.01\,\mu{\rm K}^2)\, (l/100)^{-0.42}$.
Error bars on the dust model spectrum indicate approximate sampling
variance uncertainties; although not shown here, sampling variance 
on the $r=0.2$ model is comparable in magnitude.
}
\end{centering}
\end{figure}

To illustrate how models with polarized dust and $r=0$ are 
able to fit the BICEP2 data as well as or better than models with 
negligible dust polarization and $r=0.2$, we compare the 
$B$ mode spectra for these models in Figure~\ref{fig_bb}.
Although the first bandpower of BICEP2 is low relative to the 
$r=0$ model with dust, 
the fit nevertheless remains acceptable due to considerable uncertainty from 
sampling variance in the BICEP2 field, which has an effective area
of about $380\,{\rm deg}^2$ or approximately $0.9\%$ of the sky.

\begin{figure}[t]
\begin{centering}
\includegraphics[width=5.in]{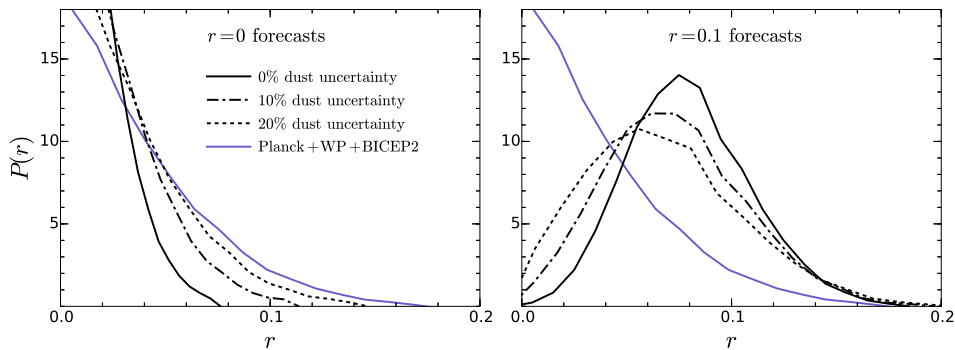}
\caption{\label{fig_1d_forecasts}
\small Forecasts for marginalized constraints on $r$ from 
Planck+WP+BICEP2 assuming lensing plus a gravity wave component
with either $r=0$ (left) or $r=0.1$ (right), with the remainder of the 
signal measured by BICEP2 coming from polarized dust. For the black curves,
this polarized dust contribution is assumed to be known within the 
BICEP2 field perfectly (solid curve), with 10\% accuracy (dot-dashed),
or with 20\% accuracy (dotted). The solid blue curve shows the
marginalized-dust constraint from Figure~\ref{fig_1d} again for comparison.
}
\end{centering}
\end{figure}

We next consider how accurate measurements of the dust polarization pattern in 
the BICEP2 field could further tighten limits on $r$. First, 
we assume that $r=0$ so that the measured BICEP2 bandpowers are 
the sum of the lensing $B$ mode power spectrum and a dust component, 
which is independently measured with a fractional uncertainty 
$\sigma_{\rm dust}$. In the most optimistic scenario where the 
dust polarization in the BICEP2 field is measured perfectly, the 
joint constraint on $r$ from Planck+WP+BICEP2 
improves significantly to $r<0.05$ (left panel of 
Figure~\ref{fig_1d_forecasts}).
If the dust polarization can be measured with $10\%$ uncertainty, 
the limit degrades to $r<0.07$; with $20\%$ uncertainty, 
$r < 0.09$; and with $30\%$ uncertainty, $r<0.10$, in which case
the direct measurement of the dust polarization provides almost 
no additional information about the tensor-to-scalar ratio.

If we perform the same analysis assuming that the true 
tensor-to-scalar ratio is $r=0.1$ (so that approximately half of
the $B$ mode power measured by BICEP2 is attributed to dust polarization, after
subtracting lensing), the Planck+WP+BICEP2 constraint is
$r = 0.08 \pm 0.03$ (right panel of Figure~\ref{fig_1d_forecasts}). 
This is biased low relative to the assumed value 
of $r$ because of the upper limit on $r$ from Planck data.
In this case, the BICEP2 data by themselves disfavor a model without
gravity waves by $\Delta \chi^2 \approx 11$ relative to the
hypothesis of $r=0.1$ plus dust.
Including uncertainty on the measured dust spectrum increases the
uncertainty on $r$ to some degree and also increases the bias in the
combined constraints; e.g., with 20\% uncertainty on dust polarization,
$r = 0.06_{-0.03}^{+0.04}$ from Planck+WP+BICEP2 and
$\Delta \chi^2 \approx 7$ in favor
of $r=0.1$ over $r=0$ using only the BICEP2 likelihood, while a 30\% 
uncertainty gives $\Delta \chi^2 \approx 5$.

\section{Frequency dependence}

One can ask whether 
there is other evidence in the current data sets that would favor $r > 0$. One argument presented by the
BICEP2 team is that the 
spectral color of the signal is inconsistent with either a pure dust or pure synchrotron 
signal at about $2.2\,\sigma$. However, the analysis was done using the first 5
bandpowers, and in all but the lowest of these bandpowers the contribution 
of the lensing spectrum, which has the spectral color of the CMB, is 
also important; this complication was not discussed in 
the BICEP2 papers.\footnote{After we posted the preprint of our paper, 
the BICEP2 team revised their frequency analysis to account for the 
lensing component \cite{2014PhRvL.112x1101A}.}
Moreover, there is likely to be both synchrotron and dust in this field and, according to recent Planck 
results, the two are strongly correlated in polarization \cite{2014arXiv1405.0871P}. While 
in typical regions the synchrotron and dust components
are roughly equal around 60\,GHz, for this field where the dust levels
are particularly low this point could 
be pushed to higher frequencies.
Because the two foregrounds are correlated and may both be present in this regime, the two
can combine into a spectral color 
that can be close to that of the pure CMB, even though separately they each have a very different spectral 
slope. 

To make these arguments more quantitive, we perform an $l$-dependent analysis. We  combine the 
the BICEP2 auto-correlation errors with the BICEP1 100\,GHz -- BICEP2 150\,GHz cross-correlation errors, 
and first assume that auto-correlation measures exactly  
the same field as the BICEP1--BICEP2 cross-correlation 
so that there is no additional sampling variance. We test two models: first, the null hypothesis with $r=0$, assuming that
the signal consists of dust scaling as $\nu^{2.3}$ relative to the
CMB,\footnote{For BICEP2, the spectral index of the CMB is $\beta=-0.7$
\cite{2014PhRvL.112x1101A}, and Planck data give a spectral index of
$\beta\approx 1.6$ for dust \cite{2013arXiv1312.1300A}, 
so the relative index is $\approx 2.3$.} and the expected lensing contribution (where 
we ignore sampling variance fluctuations). 
We could also add synchrotron at the expected level \cite{2014arXiv1404.5323F}, partially correlated
with dust \cite{2014arXiv1405.0871P}, although we note that this does not change the results much as we will show below. 
The second model is the no dust, pure CMB, hypothesis, which has the CMB color. 
The $l=45$ point measures $0.0086\,\mu{\rm K}^2$ in BICEP2 auto-power, 
which the null hypothesis predicts should be $0.0038\,\mu{\rm K}^2$ for                  
the cross-correlation with 100\,GHz BICEP1, while the pure CMB no dust
hypothesis predicts that this value is unchanged.
The BICEP2--BICEP1 100\,GHz cross-power measures $(-0.0034\pm 0.0056)\,\mu{\rm K}^2$, so the null hypothesis is favored relative to 
the no dust hypothesis by $\Delta \chi^2 \sim -3$.  For the $l=75$
point, the BICEP2 auto-power is $0.013\,\mu{\rm K}^2$, which the null
hypothesis predicts should be $0.0062\,\mu{\rm K}^2$ for 
the cross-correlation with 100\,GHz BICEP1. The measurement is $(0.025\pm
0.0065)\,\mu{\rm K}^2$, so this point favors the CMB 
hypothesis against the null hypothesis by $\Delta \chi^2 \sim 5$. We can repeat the same at the higher $l$ points, 
where the differences between the two hypotheses are smaller due to increasing lensing contribution and the errors are
larger, 
finding that overall the difference in total $\chi^2$ is 2 in favor of the no dust model. 
We find that the first two bandpowers have the most discriminatory power between 
the two hypotheses, but they nearly cancel each other out given the observed values. 
Since adding synchrotron only makes the spectral
dependence of dust+synchrotron closer to the CMB, this difference in $\chi^2$ is reduced if we include a small 
amount of correlated synchrotron polarization in our analysis. 

Another important consideration is sampling variance, which we have ignored so
far;  
we assumed that the BICEP1--BICEP2 cross-correlation analysis samples exactly the same modes as the
BICEP2 auto-correlation, but in reality this is unlikely to be the case given the differences in sky coverage and 
noise properties of the two experiments \cite{2014ApJ...783...67B,2014arXiv1403.3985B}, and
the two could in fact be measuring different realizations of dust and CMB $B$ modes. 
The large $\chi^2$ we found in the comparison between the two measurements above for 
either of the two models, especially for the first two bandpowers,
suggests that there may be additional sampling variance errors.
This would not be surprising: the sampling variance errors
for BICEP2 auto-power 
are very large, about 60\% for the first bandpower and 35\% for the second
bandpower \cite{2014arXiv1403.3985B} (see Figure~\ref{fig_bb}).\footnote{There
is also some possible evidence of sampling variance at work in the comparison
of the BICEP2 $BB$ auto-correlation and the BICEP2--Keck Array cross-correlation
presented in \cite{2014arXiv1403.3985B}, which appear to be several $\sigma$ apart for the lowest $l$ bandpowers.}
Suppose, for example, that we need to add an additional error on the order of 30\% (corresponding to 
about $0.004\,\mu{\rm K}^2$ for the dust+lensing model) because of the 
sampling variance errors. Then the difference in $\chi^2$ between the two hypotheses is further reduced to 
1.5, and presumably to even less if synchrotron is included. This is a very different conclusion than the 
one presented in the BICEP2 paper, where it was stated that dust is disfavored at $\Delta \chi^2 \sim 5$ ($2.2\,\sigma$). Most 
of the difference can be accounted for with the lensing contribution, which we included in our analysis and which
was not included in the BICEP2 analysis. 
The current frequency information is thus unlikely 
to be a strong discriminator between the dust and no dust hypotheses. 

Future Keck Array data with much lower noise could be very powerful in
discriminating between these 
hypotheses. There are a few lessons from the exercise above that should be generally applicable. 
Because of large sampling variance, care should be taken to
minimize its error by 
using the same weights for the two frequency channels. Because of the lensing effect, and the sampling 
variance associated with it, only the lowest $l$
bandpowers contain useful information for this test. A proper statistical analysis, including 
lensing, synchrotron, and dust, and accounting for both
measurement and sampling variance errors, is needed to assess the statistical significance of 
the frequency information.

\section{Discussion}

In this work we have performed a joint analysis of Planck and BICEP2, assuming lensing, primordial gravity waves, and 
dust polarization as the components contributing to the BICEP2 $B$ mode data. 
We have assumed that we do not have a reliable prior on the dust polarization in the BICEP2 field, 
so we perform a pure power spectrum analysis. 
We find that under the assumption of a dust polarization power spectrum
with free amplitude but known scaling with $l$, the  BICEP2 data prefer $r=0$.
The limits on $r$ change from $r<0.13$ with Planck to $r<0.11$ with Planck+BICEP2 (95\% c.l.) when using 
all 9 bandpowers, but remain at $r<0.13$ when only the first 5 bandpowers are
used. 
Including the latest Planck constraints on dust polarization in the BICEP2 field
from \cite{2014arXiv1409.5738P} reduces the limit to $r<0.09$.
With our analysis, we find that $V=M^2\phi^2/2$ inflation predicting $r=0.14$
is disfavored at about $2.5\,\sigma$ when the dust amplitude is unconstrained,
and at nearly $3\,\sigma$ when we apply the Planck dust polarization
constraint.
This result does not automatically 
mean that BICEP2 has no evidence for primordial gravitational waves in its data. It does, however, mean 
that the case strongly relies on careful characterization of the actual 
dust polarization contribution in the BICEP2 field, which appears to be
higher than the various estimates presented by the BICEP2 team. 
It is thus too early to celebrate the BICEP2 
results as a definitive proof of inflation. 

Moving forward, Planck dust polarization data in the BICEP2 field has
already begun to clarify the situation. However, using 
the data without processing them through the BICEP2 pipeline induces
sampling variance errors; due to various 
projections of BICEP2 data, variable noise,
and potentially highly variable dust polarization in the field, one needs to 
process Planck dust data through the BICEP2 pipeline to minimize the sampling variance. 
We note that the sampling variance errors in the first 3 BICEP2 bins, 
which have the greatest sensitivity to gravity wave $B$ modes, are around 60\%,
35\%, and 30\%. 
Even if the Planck analysis had measured low dust polarization on average 
in the 
BICEP2 field, given the potentially large sampling fluctuations within the field 
there would be no guarantee that dust contamination was actually low for the specific mode 
and area weighting used by BICEP2, unless the whole field has low dust polarization. 

For a more reliable answer and improved constraints,
a combined analysis of BICEP2 and Planck is thus required. 
We find that in the most optimistic case where the dust foreground is known perfectly and assumed to 
be consistent with the observed signal, such a combined analysis could improve 
the limits to $r<0.05$ (95\%\,c.l.), or detect a gravity wave signal with $r\sim 0.1$ 
with an uncertainty of about $0.03$ (68\%\,c.l.). 
However, various 
systematics such as bandpass and calibration leakage  may actually make this more difficult than expected. 
In addition, there is some uncertainty in the scaling of the signal from high to low frequencies, since 
the slope can vary by about 10\% \cite{2014arXiv1405.0874P}, and there is also the possibility of
the dust signal decorrelating 
between frequencies due to incoherent mixing of different dust sources. 
We show that the limits quickly degrade as we move away from perfect knowledge of the dust 
foreground. If the dust polarization can be subtracted off with 10\% residual error, then the limit becomes 
$r<0.07$; a 20\% error gives $r<0.09$ and a 30\% error gives $r<0.10$, only a minor 
improvement over the current limits presented here. 
This does not mean that such an analysis will not be useful even if the errors cannot be 
reduced below 20\%: 
while we have framed our tests conservatively in terms of 
upper limits on $r$, it is of course possible that the measured dust polarization will be below the 
BICEP2 observations, in which case there will be a residual signal in the data 
that can be attributed to primordial gravity waves. If so, the limits become detection levels and the 
importance of the dust foreground is reduced. 
We note that the lower detector noise error expected from the Keck Array
in itself does little to improve 
the constraints in the absence of a better understanding of the dust contribution. 
Another possible approach is to weight the data by the dust polarization
and eliminate patches with a high dust polarization signal. 
The BICEP/Keck Array field is relatively large and there may be regions within 
it that have lower dust polarization than average. Focusing on these could be more beneficial than 
estimating the dust contamination level for the whole observing field,
especially for 
Keck Array data with its low detector noise, but the price one has to pay is increased $E$ to $B$ mixing.

We analyzed the frequency dependence of the signal and showed that the current data do not strongly
favor the hypothesis of no foregrounds against the null hypothesis of dust and
lensing: we find that the difference in $\chi^2$
between the two hypotheses is less than 2 (i.e. less than $1.4\,\sigma$) and that 
the first bandpower, which has the largest discriminating power since it has the lowest lensing contribution,
actually favors the dust interpretation, while the second bandpower suggests just the opposite. 
Adding a possible correlated synchrotron component, which makes the spectral color closer
to the CMB, results in even 
less difference between the no foreground and foreground+lensing models. 
Future data, such as 100\,GHz measurements
from the Keck Array, will greatly improve upon this test. 
Synchrotron may complicate the situation to some extent and the level of its correlation with 
dust in the low intensity regions needs to be quantified better, but overall synchrotron does not 
appear to be a major issue. The multi-frequency information may thus be the best path forward, 
but future analyses must attempt to minimze the sampling variance errors, which can quickly dominate 
the error budget otherwise. 

What are the prospects for improving limits on $r$ from $B$ mode polarization in the future? 
The high level of dust, even in the cleanest patches of the sky
presented in \cite{Aumont}, as well as an average polarization fraction of 8--10\% in 
patches of low dust intensity \cite{2014arXiv1405.0871P}, is relatively bad news 
for the field: it means more work will be needed to separate the CMB from the
polarized dust. Multi-frequency observations will be required to 
make any such separation possible and future analyses should attempt to minimize 
sampling variance errors, which can quickly degrade the constraints. On the positive side, we have no evidence 
that we need to model more than three components in polarization, and possibly only two will be needed, 
so with good frequency coverage and low noise there are no obvious obstacles. 
There may also be small patches of the sky which are quite clean, and future surveys should 
choose carefully the regions of the sky they observe. 
Overall, we remain optimistic that future $B$ mode polarization searches will provide powerful constraints on 
inflationary models. 

M.M. and U.S. are supported in part by the NASA ATP grant NNX12AG71G.
We thank R. Flauger, W. Holzapfel, A. Lee, L. Senatore, B. Sherwin, M. White,
and M. Zaldarriaga for useful discussions. 

\bibliography{cosmo,cosmo_preprints}

\begin{thebibliography}{10}
\providecommand*{\bibinfo}[2]{#2}
\providecommand*{\eprint}[1]{#1}
\providecommand*{\url}[1]{#1}
\bibitem{2013arXiv1303.5082P}
\bibinfo{author}{{Planck Collaboration}}, \bibinfo{author}{P.~A.~R. {Ade}},
  \bibinfo{author}{N.~{Aghanim}}, \bibinfo{author}{C.~{Armitage-Caplan}},
  \bibinfo{author}{M.~{Arnaud}}, \bibinfo{author}{M.~{Ashdown}},
  \bibinfo{author}{F.~{Atrio-Barandela}}, \bibinfo{author}{J.~{Aumont}},
  \bibinfo{author}{C.~{Baccigalupi}}, \bibinfo{author}{A.~J. {Banday}},
  \emph{et~al.}, \bibinfo{journal}{ArXiv e-prints}  (\bibinfo{date}{Mar.
  2013}), \eprint{1303.5082}.
\bibitem{1997ApJ...482....6S}
\bibinfo{author}{U.~{Seljak}}, \bibinfo{journal}{\apj}
  \bibinfo{volume}{\textbf{482}}, \bibinfo{pages}{6+} (\bibinfo{date}{Jun.
  1997}).
\bibitem{1997PhRvD..55.1830Z}
\bibinfo{author}{M.~{Zaldarriaga}} and \bibinfo{author}{U.~{Seljak}},
  \bibinfo{journal}{\prd} \bibinfo{volume}{\textbf{55}}, \bibinfo{pages}{1830}
  (\bibinfo{date}{Feb. 1997}).
\bibitem{1997PhRvD..55.7368K}
\bibinfo{author}{M.~{Kamionkowski}}, \bibinfo{author}{A.~{Kosowsky}}, and
  \bibinfo{author}{A.~{Stebbins}}, \bibinfo{journal}{\prd}
  \bibinfo{volume}{\textbf{55}}, \bibinfo{pages}{7368} (\bibinfo{date}{Jun.
  1997}).
\bibitem{1998PhRvD..58b3003Z}
\bibinfo{author}{M.~{Zaldarriaga}} and \bibinfo{author}{U.~{Seljak}},
  \bibinfo{journal}{\prd} \bibinfo{volume}{\textbf{58}}, \bibinfo{pages}{23003}
  (\bibinfo{date}{Jul. 1998}).
\bibitem{2014arXiv1403.3985B}
\bibinfo{author}{{BICEP2 Collaboration}}, \bibinfo{author}{P.~A.~R. {Ade}},
  \bibinfo{author}{R.~W. {Aikin}}, \bibinfo{author}{D.~{Barkats}},
  \bibinfo{author}{S.~J. {Benton}}, \bibinfo{author}{C.~A. {Bischoff}},
  \bibinfo{author}{J.~J. {Bock}}, \bibinfo{author}{J.~A. {Brevik}},
  \bibinfo{author}{I.~{Buder}}, \bibinfo{author}{E.~{Bullock}}, \emph{et~al.},
  \bibinfo{journal}{ArXiv e-prints}  (\bibinfo{date}{Mar. 2014}),
  \eprint{1403.3985v2}.
\bibitem{2014arXiv1405.0871P}
\bibinfo{author}{{Planck Collaboration}}, \bibinfo{author}{P.~A.~R. {Ade}},
  \bibinfo{author}{N.~{Aghanim}}, \bibinfo{author}{D.~{Alina}},
  \bibinfo{author}{M.~I.~R. {Alves}}, \bibinfo{author}{C.~{Armitage-Caplan}},
  \bibinfo{author}{M.~{Arnaud}}, \bibinfo{author}{D.~{Arzoumanian}},
  \bibinfo{author}{M.~{Ashdown}}, \bibinfo{author}{F.~{Atrio-Barandela}},
  \emph{et~al.}, \bibinfo{journal}{ArXiv e-prints}  (\bibinfo{date}{May 2014}),
  \eprint{1405.0871}.
\bibitem{2014arXiv1409.5738P}
\bibinfo{author}{{Planck Collaboration}}, \bibinfo{author}{R.~{Adam}},
  \bibinfo{author}{P.~A.~R. {Ade}}, \bibinfo{author}{N.~{Aghanim}},
  \bibinfo{author}{M.~{Arnaud}}, \bibinfo{author}{J.~{Aumont}},
  \bibinfo{author}{C.~{Baccigalupi}}, \bibinfo{author}{A.~J. {Banday}},
  \bibinfo{author}{R.~B. {Barreiro}}, \bibinfo{author}{J.~G. {Bartlett}},
  \emph{et~al.}, \bibinfo{journal}{ArXiv e-prints}  (\bibinfo{date}{Sep.
  2014}), \eprint{1409.5738}.
\bibitem{Bernard}
\bibinfo{author}{{Planck collaboration}} and \bibinfo{author}{J.-P. {Bernard}},
  \bibinfo{journal}{ESLAB conference}  (\bibinfo{date}{2013}),
  \eprint{http://www.rssd.esa.int}.
\bibitem{2014PhRvL.112x1101A}
\bibinfo{author}{P.~A.~R. {Ade}}, \bibinfo{author}{R.~W. {Aikin}},
  \bibinfo{author}{D.~{Barkats}}, \bibinfo{author}{S.~J. {Benton}},
  \bibinfo{author}{C.~A. {Bischoff}}, \bibinfo{author}{J.~J. {Bock}},
  \bibinfo{author}{J.~A. {Brevik}}, \bibinfo{author}{I.~{Buder}},
  \bibinfo{author}{E.~{Bullock}}, \bibinfo{author}{C.~D. {Dowell}},
  \emph{et~al.}, \bibinfo{journal}{Physical Review Letters}
  \bibinfo{volume}{\textbf{112}}(24), \bibinfo{pages}{241101},
  \bibinfo{eid}{241101} (\bibinfo{date}{Jun. 2014}), \eprint{1403.3985}.
\bibitem{Aumont}
\bibinfo{author}{{Planck collaboration}} and \bibinfo{author}{J.~{Aumont}},
  \bibinfo{journal}{ESLAB conference}  (\bibinfo{date}{2013}),
  \eprint{http://www.rssd.esa.int}.
\bibitem{2014arXiv1405.0874P}
\bibinfo{author}{{Planck Collaboration}}, \bibinfo{author}{P.~A.~R. {Ade}},
  \bibinfo{author}{M.~I.~R. {Alves}}, \bibinfo{author}{G.~{Aniano}},
  \bibinfo{author}{C.~{Armitage-Caplan}}, \bibinfo{author}{M.~{Arnaud}},
  \bibinfo{author}{F.~{Atrio-Barandela}}, \bibinfo{author}{J.~{Aumont}},
  \bibinfo{author}{C.~{Baccigalupi}}, \bibinfo{author}{A.~J. {Banday}},
  \emph{et~al.}, \bibinfo{journal}{ArXiv e-prints}  (\bibinfo{date}{May 2014}),
  \eprint{1405.0874}.
\bibitem{Flauger}
\bibinfo{author}{R.~{Flauger}}, \bibinfo{author}{J.~C. {Hill}}, and
  \bibinfo{author}{D.~N. {Spergel}}, \bibinfo{journal}{\jcap}
  \bibinfo{volume}{\textbf{8}}, \bibinfo{pages}{39}, \bibinfo{eid}{039}
  (\bibinfo{date}{Aug. 2014}), \eprint{1405.7351}.
\bibitem{2013arXiv1303.5076P}
\bibinfo{author}{{Planck Collaboration}}, \bibinfo{author}{P.~A.~R. {Ade}},
  \bibinfo{author}{N.~{Aghanim}}, \bibinfo{author}{C.~{Armitage-Caplan}},
  \bibinfo{author}{M.~{Arnaud}}, \bibinfo{author}{M.~{Ashdown}},
  \bibinfo{author}{F.~{Atrio-Barandela}}, \bibinfo{author}{J.~{Aumont}},
  \bibinfo{author}{C.~{Baccigalupi}}, \bibinfo{author}{A.~J. {Banday}},
  \emph{et~al.}, \bibinfo{journal}{ArXiv e-prints}  (\bibinfo{date}{Mar.
  2013}), \eprint{1303.5076}.
\bibitem{2013arXiv1312.3313S}
\bibinfo{author}{D.~{Spergel}}, \bibinfo{author}{R.~{Flauger}}, and
  \bibinfo{author}{R.~{Hlozek}}, \bibinfo{journal}{ArXiv e-prints}
  (\bibinfo{date}{Dec. 2013}), \eprint{1312.3313}.
\bibitem{2000ApJ...538..473L}
\bibinfo{author}{A.~{Lewis}}, \bibinfo{author}{A.~{Challinor}}, and
  \bibinfo{author}{A.~{Lasenby}}, \bibinfo{journal}{\apj}
  \bibinfo{volume}{\textbf{538}}, \bibinfo{pages}{473} (\bibinfo{date}{Aug.
  2000}).
\bibitem{2002PhRvD..66j3511L}
\bibinfo{author}{A.~{Lewis}} and \bibinfo{author}{S.~{Bridle}},
  \bibinfo{journal}{\prd} \bibinfo{volume}{\textbf{66}}(10),
  \bibinfo{pages}{103511} (\bibinfo{date}{Nov. 2002}),
  \eprint{arXiv:astro-ph/0205436}.
\bibitem{2014arXiv1409.7025C}
\bibinfo{author}{C.~{Cheng}}, \bibinfo{author}{Q.-G. {Huang}}, and
  \bibinfo{author}{S.~{Wang}}, \bibinfo{journal}{ArXiv e-prints}
  (\bibinfo{date}{Sep. 2014}), \eprint{1409.7025}.
\bibitem{2014arXiv1405.1390A}
\bibinfo{author}{B.~{Audren}}, \bibinfo{author}{D.~G. {Figueroa}}, and
  \bibinfo{author}{T.~{Tram}}, \bibinfo{journal}{ArXiv e-prints}
  (\bibinfo{date}{May 2014}), \eprint{1405.1390}.
\bibitem{2013arXiv1312.1300A}
\bibinfo{author}{A.~{Abergel}}, \bibinfo{author}{P.~A.~R. {Ade}},
  \bibinfo{author}{N.~{Aghanim}}, \bibinfo{author}{M.~I.~R. {Alves}},
  \bibinfo{author}{G.~{Aniano}}, \bibinfo{author}{C.~{Armitage-Caplan}},
  \bibinfo{author}{M.~{Arnaud}}, \bibinfo{author}{M.~{Ashdown}},
  \bibinfo{author}{F.~{Atrio-Barandela}}, \bibinfo{author}{J.~{Aumont}},
  \emph{et~al.}, \bibinfo{journal}{ArXiv e-prints}  (\bibinfo{date}{Dec.
  2013}), \eprint{1312.1300}.
\bibitem{2014arXiv1404.5323F}
\bibinfo{author}{U.~{Fuskeland}}, \bibinfo{author}{I.~K. {Wehus}},
  \bibinfo{author}{H.~K. {Eriksen}}, and \bibinfo{author}{S.~K. {N{\ae}ss}},
  \bibinfo{journal}{ArXiv e-prints}  (\bibinfo{date}{Apr. 2014}),
  \eprint{1404.5323}.
\bibitem{2014ApJ...783...67B}
\bibinfo{author}{D.~{Barkats}}, \bibinfo{author}{R.~{Aikin}},
  \bibinfo{author}{C.~{Bischoff}}, \bibinfo{author}{I.~{Buder}},
  \bibinfo{author}{J.~P. {Kaufman}}, \bibinfo{author}{B.~G. {Keating}},
  \bibinfo{author}{J.~M. {Kovac}}, \bibinfo{author}{M.~{Su}},
  \bibinfo{author}{P.~A.~R. {Ade}}, \bibinfo{author}{J.~O. {Battle}},
  \emph{et~al.}, \bibinfo{journal}{\apj} \bibinfo{volume}{\textbf{783}},
  \bibinfo{pages}{67}, \bibinfo{eid}{67} (\bibinfo{date}{Mar. 2014}).

\end{thebibliography}
\bibliographystyle{revtex}

\end{document}